\documentclass[reprint,amsmath,amssymb]{revtex4-2}
\usepackage{comment}

\usepackage{graphicx}
\usepackage{amsmath}
\usepackage{amsfonts}
\usepackage{dcolumn}
\usepackage{mathrsfs}
\usepackage{xcolor}
\usepackage{calligra}
\usepackage{bm}
\usepackage{enumerate}
\usepackage{pgfplots}
\pgfplotsset{compat=1.17}
\usepackage{tikz}
\usetikzlibrary{decorations.markings}
\usetikzlibrary{plotmarks}
\usepackage{diagbox}
\usepackage{amsthm}
\theoremstyle{remark}

\newif\iffig
\figfalse
\newcommand{\tensorchi}{\chi}
\newcommand{\tensorgamma}{\gamma}
\newcommand{\Ep}{\mathbf{E}^P}

\DeclareMathOperator{\spn}{span}

\begin{document}

\title{Synthesis of resonant modes in electromagnetics\\}

\author{Antonello Tamburrino}
\affiliation{Department of Electrical and Information Engineering \lq\lq M. Scarano\rq\rq, Universit\`a degli Studi di Cassino e del Lazio Meridionale, Via G. Di Biasio n. 43, 03043 Cassino (FR), Italy.\\
\\
Department of Electrical and Computer Engineering, Michigan State University, East Lansing, MI-48824, USA.\\
\\
e-mail: antonello.tamburrino@unicas.it}

\author{Carlo Forestiere, Giovanni Miano, Guglielmo Rubinacci}
\affiliation{Department of Electrical Engineering and Information Technology, Universit\`{a} degli Studi di Napoli Federico II, via Claudio 21,  Napoli, 80125, Italy}

\author{Salvatore Ventre}
\affiliation{Department of Electrical and Engineering Information \lq\lq M. Scarano\rq\rq, Universit\`a degli Studi di Cassino e del Lazio Meridionale, Via G. Di Biasio n. 43, 03043 Cassino (FR), Italy.}


\begin{abstract}
Resonant modes determine the response of electromagnetic devices, including dielectric and plasmonic resonators. Relying on the degrees of freedom that metamaterials provide, this contribution shows how to design, at will, the resonant modes of a dielectric object placed in an unbounded space. Specifically, the proposed method returns in analytical form the spatial distribution of the dielectric susceptibility tensor for which the object exhibits resonances  at prescribed frequencies and spatial distribution of the polarization. Together with the synthesis of the material, two key concepts are introduced: the controlled tunability of the resonant modes and the number of essential modes, i.e. the number of modes that uniquely characterize the spatial distribution of the dielectric susceptibility.
Moreover, this approach can be applied to design the resonant modes of any system where the constitutive relationship is linear and local.

\end{abstract}
\maketitle

Media with spatially inhomogeneous refractive index have fascinated the humankind for millennia, exhibiting counter-intuitive effects such as mirages, or fata morgana.  Archaeological evidence indicates that humans learned how to engineer the refractive index variations to make lenses in antiquity, spanning several millennia. More recently, nano-fabrication techniques, the discovery of materials with tunable permittivity, and the introduction of the metamaterial concept \cite{engheta_metamaterials_2006} have greatly expanded the landscape of feasible permittivity distributions for the electromagnetic design. Anisotropic and even continuous {\it effective} variations of the permittivity can be now implemented.

{
Using the degrees of freedom in the choice of the materials, it is possible to control the electromagnetic field  as shown by Pendry et al \cite{pendry_controlling_2006} by introducing trasformation optics \cite{pendry_controlling_2006,leonhardt_optical_2006}. They showed that, the  permittivity and permeability effectively determine a curved spatial geometry for the electromagnetic field.  Thus, leveraging on this analogy, they showed how the anisotropic and inhomogeneous permittivity and permeability profiles to redirect the electromagnetic field in a prescribed way. Recently, several optimization methods have been introduced to design materials to achieve a prescribed electromagnetic response, incorporating at the same time fabrication constraints  
\cite{hughes_adjoint_2018,yao_intelligent_2019}.

In this manuscript, we take a fresh path to the design of electromagnetic resonances of a scatterer, which plays a central role in electromagnetic devices, e.g. \cite{lalanne_light_2018}.  Plasmonic and dielectric nano-resonators are an interesting example. When the resonance condition is met, the near-field and far-field characteristics of the device are dominated by the corresponding resonant mode.

We introduce a theoretical framework that enables the synthesis of the spatial distribution of the permittivity profile of a dielectric object, to design its resonant modes, i.e. polarization current density distributions. The designer preliminary specifies, in the spatial domain occupied by the object, one or several modes, together with the corresponding resonant frequencies. Then, the synthesis process returns the possibly inhomogeneous and anisotropic permittivity profile which guarantees that the dielectric object exhibits the prescribed modes at the specified resonance frequencies.  It is a direct method: it does not require the use of any optimization approaches, but explicitly returns the analytical solution in a single step. The syntheses approach leverages on a formulation of the generalized eigenvalue problem where the contributions of the material and of the electromagnetic field are separated. Yet, this approach is very general: it can be applied to any system where the constitutive relationship is linear and non-spatially dispersive. For instance, it can be used  to design the  properties of an elastic material to control its vibrational modes.

In addition, the proposed framework allows one to clearly identify the physical feasibility and limitations inherent to the problem of the design of the modes.  The main outcome is that the maximum number of modes (\emph{essential modes}) that can be prescribed at a given resonance frequency, is equal to the dimension of the problem (two for a 2D problem and three for a 3D problem). These are inherent physical limits unveiled by the proposed framework.

Finally, we also address the problem of the \emph{tunability} where, by scaling the dielectric susceptibility, we can change completely the resonance property in a controlled way. This feature enables the design of tunable materials, where one can adapt the response of the material dynamically, according to specific needs.}

\section*{Modes and Eigenvalue problem}

\begin{figure}
    \centering
    \includegraphics[width=0.8\columnwidth]{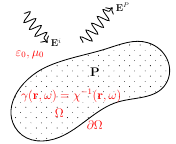}
    \caption{\textbf{Sketch of the problem.} A finite size dielectric object of inhomogeneous and anisotropic susceptibility
    $\chi \left(  \mathbf{r}, \omega \right)$ is placed into an unbounded space. The proposed method finds $\chi \left(  \mathbf{r},\omega\right)$
    such that the objects exhibits resonance modes of prescribed spatial distribution at prescribed frequencies.}
    \label{fig:Particle}
\end{figure}


We consider a linear, nonmagnetic and non-spatially dispersive dielectric of finite size, shown in Fig. \ref{fig:Particle}. We denote the space occupied by the dielectric with $\Omega$, its boundary by $\partial \Omega$, the (unit vector) normal to $\partial V$ that points outward by $\mathbf{n}$.

Under these assumptions, the polarization density $\mathbf{P}$ is given by $\mathbf{P}\left(  \mathbf{r},\omega\right)
 = \varepsilon_{0} \chi \left(  \mathbf{r},\omega\right)
\cdot\mathbf{E}\left(  \mathbf{r},\omega\right)  $, where $\tensorchi$ is
the  dielectric susceptibility tensor, $\omega$ is the angular frequency
(the $e^{j\omega t}$ time behavior is assumed), $\varepsilon_0$ is the vacuum permittivity, and $\cdot$ corresponds to the
usual dot product between tensors and vectors. 

When the dielectric scatterer is excited by an external electric field $\mathbf{E}^{i}$, the total electric field $\mathbf{E}$ can be written as the sum of $\mathbf{E}^{i}$ and of the reaction field $\mathbf{E}^{\mathtt{P}}$ due to the presence
of the polarization current density $j\omega\mathbf{P}$. The constitutive
relation can be written as
\begin{equation}
\frac{1}{\varepsilon_{0}}\tensorgamma \left( \mathbf{r}, \omega \right)
\cdot \mathbf{P} \left( \mathbf{r}, \omega \right) - \Ep \left(
\mathbf{r}, \omega \right)  = \mathbf{E}^{i}\left( \mathbf{r}, \omega \right)
\text{ in } \Omega,
\label{eq:Constitutive}
\end{equation}
where tensor $\mathbf{\tensorgamma}$ is the pointwise inverse of
$\tensorchi$, i.e. $\mathbf{\tensorgamma}\left(  \mathbf{r},\omega\right)
=\tensorchi^{-1}\left(  \mathbf{r},\omega\right)  $.



Let $\mathcal{E}\left(  \omega\right)$
be the operator giving the electric field produced by a prescribed polarization
density field $\mathbf{P}$ radiating in the free space at frequency $\omega
$ \cite{van_bladel_electromagnetic_2007}:%
\begin{equation*}
\mathbf{E}^{P}\left(  \mathbf{r}\right)  =j\omega\int_{\Omega}\mathbf{G}%
\left(  \mathbf{r-r}^{\prime}\right)  \mathbf{P}\left(  \mathbf{r}^{\prime
}\right)  \text{d}S^{\prime}%
\end{equation*}
where $\mathbf{G}$ is the proper electric-electric dyadic Green function.

For any prescribed angular frequency $\omega$, the electromagnetic scattering is
governed by the integral equation
\begin{equation}
\frac{1}{\varepsilon_{0}}\mathbf{\tensorgamma}\cdot\mathbf{P} - \mathcal{E}\left(
\omega\right)  \mathbf{P}=\mathbf{E}^{i} \qquad \text{ in }\Omega.
\label{eq02}%
\end{equation}


Two particularly significant auxiliary eigenvalue problems can be defined starting from Eq. \ref{eq02}, setting the exciting field to zero, and assigning the material tensor $\mathbf{\tensorgamma}$.

Quasi Normal Modes \cite{kristensen_modes_2013,muljarov_brillouin-wigner_2010,lalanne_light_2018,lalanne_quasinormal_2019} (QNM) are nontrivial solutions $\omega$ and $\mathbf{P}$ of
\begin{equation}
\mathcal{E}\left(  \omega\right)  \mathbf{P=}\frac{1}{\varepsilon_{0}%
}\mathbf{\tensorgamma}\cdot\mathbf{P}\text{ in }\Omega.\label{eq10}%
\end{equation}
QNM are often used to characterize micro- and nano-  resonators \cite{lalanne_quasinormal_2019}, enabling the calculation of synthetic parameters such as the quality factor, the mode volume \cite{kristensen_generalized_2012,sauvan_theory_2013}, and the Purcell factor. QNM are also used to expand the response of micro-nanoresonators by \cite{muljarov_exact_2016} highlighting the contribution of the individual modes in the overall scattering response. The eigen-frequencies $\omega$ are complex numbers, i.e. $\omega \in \mathbb{C}$, and $(\omega, \mathbf{P})$ forms a (generalized)
eigenvalue/eigenvector pair.


Material Modes are nontrivial solution $\xi \in \mathbb{C}$ and $\mathbf{P}$ of
\begin{equation}
\label{eq05}
\mathcal{E}\left(  \omega\right)  \mathbf{P}=\xi\frac{1}{\varepsilon_{0}%
}\mathbf{\tensorgamma}\cdot\mathbf{P}\text{\textbf{\ }in }\Omega,
\end{equation}
where the frequency $\omega \in \mathbb{C}$ is prescribed.
$\xi$ and $\mathbf{P}$ form a (generalized)
eigenvalue/eigenvector pair.

These modes for $\omega \in \mathbb{R}$ and uniform and isotropic 
material ($\tensorchi(\mathbf{r}) = \chi$ scalar constant in $\Omega$) have been already investigated in \cite{bergman_theory_1980,forestiere_material-independent_2016,forestiere_volume_2018,pascale_full-wave_2019}, and have been used to expand the electromagnetic response of nano-resonators \cite{forestiere_material-independent_2016,forestiere_nanoparticle_2017,pascale_full-wave_2019}, and also to design the scalar permittivity of a homogeneous object to achieve a prescribed scattering response, such as scattering cancellation or maximization \cite{forestiere_material-independent_2016,pascale_spectral_2017,forestiere_directional_2019}.


In this work $\chi$ may be non uniform and/or non isotropic, and $\omega$ may be complex. The characteristic feature of the eigenvalue/eigenvector pair for (\ref{eq05}) is to be a
\textit{homogeneous} function of $\tensorchi$, i.e. if $\tensorchi^{\prime}=\alpha\tensorchi$ then%
\begin{equation*}
\mathbf{P}^{\prime}  =\mathbf{P}; \qquad 
\frac{1}{\xi^{\prime}} =\alpha\frac{1}{\xi}
\end{equation*}
is an eigenvalue/eigenvector pair for $\tensorchi^{\prime}$. Specifically, the eigenvector $\mathbf{P}$ is a 0-degree homogeneous function, whereas the reciprocal of the eigenvalue $\xi$ is a $1-$degree homogeneous function.
After this property, we term these modes as Homogeneous Material Modes. Homogeneous Material Modes have been successfully introduced in low-frequency electromagnetism for eddy current tomography \cite{su_monotonicity_2017,tamburrino_monotonicity_2021}.

{
A unique feature of Material Modes and, more in general, of Homogeneous Material Modes, is that since the eigenvalue $\xi$ and the eigenvector are homogeneous function of $\chi$, it is possible to tune on different resonant modes the electromagnetic system by scaling the susceptibility. This feature, which we call \lq\lq tunabilty\rq\rq, opens the door to a systematic design of reconfigurable materials and will be discussed in detail in a subsequent Section.
}

\section*{Synthesis of Modes (SOM)}
\label{sec:SOM}
In this Section, we introduce a theoretical framework enabling the synthesis of the dielectric permittivity tensor $\tensorchi=\tensorchi%
\left(  \mathbf{r}, \omega \right)$ of the object, such that it exhibits the set of resonance modes $\left\{  \left(\omega_{k},\xi_{k},\mathbf{P}_{k}\right) \right\}_{k=1\ldots N}$ at prescribed frequencies $\omega_k$. Each individual mode is described by the triplet $\left(  \omega_{k},\xi_{k},\mathbf{P}_{k}\right)$. Hereafter, $\omega_k$ is referred as the frequency eigenvalue, $\xi_k$ as the material eigenvalue, and $\mathbf{P}_{k}$ as the spatial mode. The problem consists in solving for a proper $\gamma_k \left( \mathbf{r} \right) = \gamma \left( \mathbf{r}, \omega_k \right)$, the set of equations imposing the modes
\begin{equation}
\label{eq:ModeEq}
\mathcal{E} \left( \omega_k \right) \mathbf{P}_k=\xi_k \frac{1}{\varepsilon_{0}} \mathbf{\tensorgamma}_k \cdot \mathbf{P}_k \text{\textbf{\ }in } \Omega \text{, for } k=1, \ldots, N.
\end{equation}

The synthesis is carried out in two steps. First, we solve the problem at each prescribed angular frequency $\omega_{k}$, by evaluating $\gamma_{k}$, as solution of \eqref{eq:ModeEq}. Then, we interpolate in the frequency domain the collection of tensors $\chi_1, \ldots, \chi_{N}$, being $\chi_k = \gamma_k^{-1}$

Hereafter, we consider the $\texttt{TE}_z$ scenario where the electromagnetic problem is $x_{3}-$ invariant and
the electric field is transverse to the $x_{3}-$axis. This is a 2D case where
the tensor $\tensorchi$ is of the type $\tensorchi\left(  \mathbf{r}, \omega \right)  =\sum_{l,m=1}^{2}\chi_{lm}\left(  \mathbf{r}, \omega \right)  \mathbf{e}%
_{l} \, \mathbf{e}_{m}$, the electric field is $\mathbf{E}\left(  \mathbf{r}, \omega
\right)  =E_{1}\left(  \mathbf{r}, \omega \right)  \mathbf{e}_{1}+E_{2}\left(
\mathbf{r}, \omega \right)  \mathbf{e}_{2}$, $\mathbf{r=}x_{1}\mathbf{e}_{1}%
+x_{2}\mathbf{e}_{2}$ and $\mathbf{e}_{1}$ and $\mathbf{e}_{2}$ are the unit
vectors along the $x_{1}$ and $x_{2}$ directions, respectively. The elements of the Green function are given in Appendix \ref{sec:Green}.

\subsection*{Synthesis of Modes at a prescribed angular frequency}
\label{SOM_PAF}

Given a prescribed angular frequency $\omega_{k}$, we distinguish two cases:
(i) a single mode is prescribed or (ii) two modes are prescribed. In a 3D setting, one have to include also the third case when three modes are prescribed. The treatment of this case is nothing but a straightforward extension of the one needed when two modes are prescribed.

{ \it Single mode case\label{OneMode}.} Let $\left(  \omega_{k},\xi_{k},\mathbf{P}_{k}\right)$ be an individual prescribed resonances modes at frequency $\omega_{k}$, where $\omega_{j}%
\neq\omega_{k}$ for $j\neq k$. The solution of equation (\ref{eq05}) can
be expressed in explicit form as%
\begin{equation}
\mathbf{\tensorgamma}_{k} \left(  \mathbf{r} \right) = \frac{\varepsilon_{0}%
\Ep_{k} \left(  \mathbf{r}\right)  }{\xi_{k}\left\vert
\mathbf{P}_{k}\left(  \mathbf{r}\right)  \right\vert ^{2}}\mathbf{P}_{k}%
^{\ast}\left(  \mathbf{r}\right)  +\alpha_{k}\left(  \mathbf{r}\right)
\mathbf{v}_{k}\left(  \mathbf{r}\right)  \mathbf{p}_{k}^{\ast}\left(
\mathbf{r}\right),
\label{eq:solution}
\end{equation}
where $\ast$ is the complex conjugate operation, $\Ep_k=\mathcal{E}\left(  \omega_{k}\right)  \mathbf{P}_{k}$, $\mathbf{p}%
_{k}\left(  \mathbf{r}\right)  \perp \mathbf{P}_{k}\left(  \mathbf{r}\right)  $
for almost everywhere (a.e.) $\mathbf{r}\in\Omega$ \footnote{Here $\mathbf{a}\left(  \mathbf{r}%
\right)  \bot\mathbf{b}\left(  \mathbf{r}\right)  $ means that $\mathbf{a}%
^{\ast}\left(  \mathbf{r}\right)  \cdot\mathbf{b}\left(  \mathbf{r}\right)
=0$.}, $\mathbf{v}_{k}$ is an arbitrary vector field and $\alpha_{k}$ is an
arbitrary scalar field. The solution $\gamma_k$ given in equation (\ref{eq:solution}) can be easily verified by
plugging it in equation (\ref{eq:ModeEq}).

Finally, we highlight that by means of the explicit solution of equation (\ref{eq:solution}) one can easily check if
$\tensorgamma_{k}$ is bounded or continuous. Specifically, we have that if $\Ep_{k}$ and $\mathbf{P}_{k}$ are continuous (piecewise continuous) and
$\left\vert \Ep_{k}\right\vert \mathbf{/}\left\vert \mathbf{P}%
_{k}\right\vert $ is bounded, then $\mathbf{\tensorgamma}_{k}$ is continuous (piecewise continuous).

{\it Two isofrequential modes.}
\label{TwoModes}
Let $\omega_{1}=\omega_{2}\neq\omega_{j}$ for $j>2$, and $\left(  \omega_{1},\xi_{1},\mathbf{P}_{1}\right)  $ and $\left(
\omega_{2},\xi_{2},\mathbf{P}_{2}\right)  $ be the prescribed resonances modes. Let the solution be expressed as%
\begin{equation}
\mathbf{\tensorgamma}_1 \left(  \mathbf{r} \right)= \sum_{l,m=1}^{2} \Gamma_{lm} \left( \mathbf{r} \right) \mathbf{U}_{l}\left( \mathbf{r} \right) \mathbf{P}_{m}^{\ast}\left( \mathbf{r} \right).
\label{eq:solution-two}
\end{equation}
where $\Gamma_{lm} \left( \mathbf{r} \right) \in \mathbb{C}$ and
\[
\mathbf{U}_{l} = \frac{\varepsilon_{0} \mathcal{E} (\omega_1) \mathbf{P}_{l}}{\xi_{l}},\ l=1,2.
\]
To find the unknown coefficients $\Gamma_{lm}$, we observe that by imposing Eq. \eqref{eq:ModeEq} on the two prescribed resonance modes we have:
\begin{equation}
    \mathbf{U}_r \left( \mathbf{r} \right)= \gamma_1 \left( \mathbf{r} \right) \cdot \mathbf{P}_r \left( \mathbf{r} \right) \text{ for a.e. } \mathbf{r} \in \Omega, \text{ and } r=1,2.
\end{equation}
Then, by left multiplying this expression by $\mathbf{U}^\ast_s\left( \mathbf{r} \right)$, we have
\[
\mathbf{U}_{s}^{\ast}\cdot\mathbf{U}_{t}=\sum_{l,m=1}^{2}\left(
\mathbf{U}_{s}^{\ast}\cdot\mathbf{U}_{l}\right)  \Gamma_{lm}\left(
\mathbf{P}_{m}^{\ast}\cdot\mathbf{P}_{t}\right)  \text{ in }\Omega,\ s,t=1,2,
\]
which, in matrix form, gives
\begin{equation}
\mathbf{G}_{U} \left( \mathbf{r} \right) = \mathbf{G}_{U} \left( \mathbf{r} \right) \mathbf{\Gamma} \left( \mathbf{r} \right) \mathbf{G}_{P} \left( \mathbf{r} \right),
\label{eq12}%
\end{equation}
where $\left(  G_{U}\right)  _{st}=\mathbf{U}_{s}^{\ast}\cdot\mathbf{U}_{t}$,
$\left(  G_{P}\right)  _{ik}=\mathbf{P}_{i}^{\ast}\cdot\mathbf{P}_{k}$ and
$\mathbf{\Gamma}$ is the matrix made by the unknown coefficients $\Gamma_{lm}$.

When both $\mathbf{G}_{U}$ and $\mathbf{G}_{P}$ are invertible at location $\mathbf{r}$, the solution of (\ref{eq12}) exists, is unique and is given by
\begin{equation}
\label{Gamma2D}
\mathbf{\Gamma} \left( \mathbf{r} \right) = \mathbf{G}_{P}^{-1} \left( \mathbf{r} \right).
\end{equation}
In the remaining cases, i.e. $\mathbf{G}_{P}$ and/or $\mathbf{G}_{U}$ non invertible, the solution may not exist or be unique.

It is worth noting that matrices $\mathbf{G}_{U}$ and $\mathbf{G}_{P}$ are Gram matrices and, therefore, $\mathbf{G}_{U}\mathbf{=G}_{U}^{\dag}$, $\mathbf{G}%
_{U}\mathbf{\geq 0}$, $\mathbf{G}_{P}\mathbf{=G}_{P}^{\dag}$ and
$\mathbf{G}_{P}\mathbf{\geq 0}$.

\subsection*{Parameterization of the frequency response}

Once the inverse of the susceptibility tensor is found at each each prescribed angular frequency
$\omega_k$, we need to reconstruct the dispersion relation $\tensorchi\left(  \mathbf{r},\omega\right)$, which has to satisfy the causality throught the Kramers-Kronig conditions and the Hermitian symmetry, namely $ \tensorchi \left( \mathbf{r},-\omega \right)=\tensorchi^* \left( \mathbf{r},\omega \right)$. To this purpose, we parameterize the dispersion relation, as follows
\begin{equation}
\tensorchi\left(  \mathbf{r},\omega\right)  =\sum_{m=1}^{M}\mathbf{a}%
_{m}\left(  \mathbf{r}\right)  \varphi_{m}\left(  \omega\right)  
\label{eq:DielectricSusceptibility}%
\end{equation}
where $M$ is the number of terms, each expansion function $\varphi_{m}$ is causal and Hermitian and each tensor field $\mathbf{a}_m$ is real. The $\varphi_{m}$s depend on the actual realization of the artificial material. A possible choice consists in assuming each expansion function $\varphi_{m}$ of the
Lorentz-Drude type:
\begin{equation}
\varphi_{m}\left(  \omega\right)  =\frac{\omega_{p,m}^{2}}{\left(  \omega_{0,m}
^{2}-\omega^{2}\right)  +j\omega\beta_{m}},
\label{eq:Drude}
\end{equation}
where causality requires $\beta_{m}>0$.

Tensors fields $\mathbf{a}_{m}$s can be found by point matching, for instance. Within this approach, we enforce the following constraints $\forall k =1, \dots, N$
\begin{align}
 \label{eq:PointMatchingA}
\sum_{m=1}^{M}\mathbf{a}_{m} (\mathbf{r}) \operatorname{Re}\{\varphi_{m}\left(  \omega
_{i}\right)\}  & = \operatorname{Re} \{ \gamma_k^{-1} (\mathbf{r}) \}
,\ \\
\sum_{m=1}^{M}\mathbf{a}_{m} (\mathbf{r}) \operatorname{Im}\{\varphi_{m}\left(  \omega
_{i}\right)\}  & = \operatorname{Im} \{ \gamma_m^{-1} (\mathbf{r}) \}
,\ 
\label{eq:PointMatchingB}
\end{align}
where $\operatorname{Re} \{\cdot \}$ and $\operatorname{Im} \{\cdot \}$ are the real and imaginary parts of their argument, respectively. Moreover, from \eqref{eq:PointMatchingA} and \eqref{eq:PointMatchingB}, it follows that $M=2\,N$ to have existence and uniqueness of the solutions in terms of the unknown tensor fields $\mathbf{a}_m$s.

We remark that parameters $\omega_{p,m}$, $\omega_{0,m}$ and $\beta_m$ depend on the actual realization of the artificial material. For instance, $\omega_{0,m}$ does not need to be equal to the resonant (angular) frequency $\omega_m$ prescribed for the Synthesis of the Modes. In the remaining of the paper we select parameters $\omega_{p,m}$, $\omega_{0,m}$ and $\beta_m$, to avoid the appearance of any resonance due to the expansion functions, at the resonant frequencies prescribed for the Synthesis of the Modes.

\section*{Tunability and Essential Modes}
The tunability of the resonance refers to the possibility of \lq\lq changing\rq\rq \ the properties of a material in a controlled manner. The Synthesis of Modes entails tunability in a natural manner via the material eigenvalues $\xi_k$.

Indeed, after \eqref{eq:ModeEq}, we have that a material with dielectric permittivity given by $\chi / \xi_k$, being $\chi$ the result of the synthesis of modes, resonates at the angular frequency given by $\omega_k$. In other terms, we can control the frequency behaviour of a material (value of the frequency resonances and spatial distribution of the related mode), by simply scaling $\chi$ by a proper factor. 
From another perspective, the proposed approach to the synthesis of the modes allows to get the resonance frequencies and related spatial modes as a function of an individual parameter: a scaling factor in front of the synthesized $\chi$.
This feature open the door to a systematic design of reconfigurable materials.

The concept of essential modes refers to the maximum number of modes that can be arbitrarily prescribed at a given angular frequency $\omega_k$. Equation \eqref{Gamma2D} provide the values of the $\Gamma_{lm}$ giving the sought inverse of the dielectric susceptivity tensor in \eqref{eq:solution-two}. This equation shed light on a special and not obvious physical feature of the modes: \emph{two modes are capable of defining uniquely the material property of the scatterer, at the prescribed angular frequency}. In other words, \emph{$\gamma(\cdot,\omega_k)$ is in a one-to-one correspondence with two of its modes at $\omega_k$}, at $\omega_k$. From another perspective, only two modes can be assigned in a completely independent manner or, equivalently, \emph{all the modes depend upon two arbitrarily selected modes}, at a prescribed angular frequency.

We term two arbitrary modes in a one-to-one correspondence with $\chi(\cdot,\omega_k)$ as \emph{essential modes}.

It is worth noting that he number of essential modes is two in a 2D problem and three in a 3D problem.

\section*{Application of the Theory of Synthesis of Modes}

In this Section, we show the effectiveness of the resonance synthesis method by means of three application examples.  We demonstrate (i) the capability of the method to synthesise several modes, each one having prescribed polarization density distribution at prescribed frequencies, (ii) the tunability of resonant response, by a proper scaling of the dielectric susceptibility tensor and (iii) the concept of essential modes. In the first two examples, the reference geometry is an indefinite cylinder with square $L\times L$ cross-section with $L=10$cm under the \texttt{TEz} illumination. In the third example the geometry consist of coated spherical gold nanoparticle.

\begin{figure}
    \centering
    \includegraphics[width=\columnwidth]{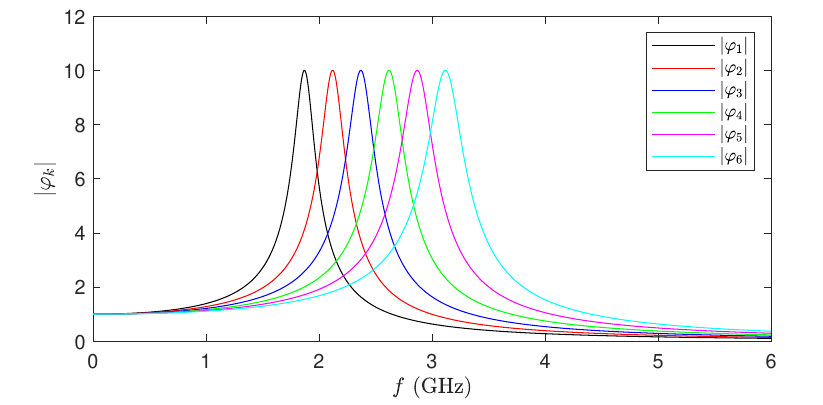}
    \caption{\textbf{Magnitude of the expansion functions.} The peaks are not able to produce significant resonances and their bandwidth provides a proper interpolation over the bandwidth of interest.}
    \label{fig:ExpFun}
\end{figure}

The numerical model for solving the \texttt{TEz} electromagnetic problem is derived from Ref. \cite{richmond_te-wave_1966}.  The parameters of the Lorentz-Drude expansion functions  $\varphi_k$, introduced in Eq. \eqref{eq:Drude}, are given in Table \ref{TB01}. The plot of each individual expansion function is shown in Figure \ref{fig:ExpFun}. The positions of the peaks of the expansion function are uniformly spaced over the bandwidth of interest. We assume $\omega_{p,k}=\omega_{0,k}$ and $\beta_k=0.1\omega_{0,k}$. With this latter choice, each expansion function is localized in a neighborhood of its peak position, but does not present a sharp resonance that could hide those arising from the Synthesis of Modes. The amplitude and the shape of the expansion function are briefly discussed in Appendix \ref{sec:LorentzDrude}.
\begin{table}[htb]
    \centering
    \begin{tabular}{c c c c c}
    \hline
    $k$ & $f_{0,k}\ $(Hz) & $\omega_{0,k}\ $(rad/s) & $\omega_{p,k}\ $(rad/s) & $\beta_{k}\ $(rad/s)\\
    \hline
    1 & $1.875\times10^{9}$ & $11.781\times10^{9}$ & $11.781\times10^{9}$ & $1.1781\times10^{9}$\\
    2 & $2.125\times10^{9}$ & $13.352\times10^{9}$ & $13.352\times10^{9}$ & $1.3352\times10^{9}$\\
    3 & $2.375\times10^{9}$ & $14.923\times10^{9}$ & $14.923\times10^{9}$ & $1.4923\times10^{9}$\\
    4 & $2.625\times10^{9}$ & $16.493\times10^{9}$ & $16.493\times10^{9}$ & $1.6493\times10^{9}$\\
    5 & $2.875\times10^{9}$ & $18.064\times10^{9}$ & $18.064\times10^{9}$ & $1.8064\times10^{9}$\\
    6 & $3.125\times10^{9}$ & $19.635\times10^{9}$ & $19.635\times10^{9}$ & $1.9635\times10^{9}$\\
    \hline
    \end{tabular}
    \caption{The parameters for the interpolating functions appearing in
equation (\ref{eq:Drude}).}
    \label{TB01}
\end{table}

{\it Synthesis of the modes.} In this first application, we prescribe the modes at the three angular frequencies shown in Table \ref{tab:TB1}. Specifically, at the angular frequency $\omega_1$ we prescribe two modes: the first one has a polarization density field $\overline{\mathbf{P}}_0$, whose shape resembles the number ``0" and it is associated with the eigenvalue $\xi_A=1$; the second mode has a polarization density field $\overline{\mathbf{P}}_{1}$, whose shape resembles the number ``1" and it is associated with the eigenvalue $\xi_B=2$. At the angular frequency $\omega_2$, we prescribe the modes $\overline{\mathbf{P}}_{1}$ and $\overline{\mathbf{P}}_{2}$, where $\overline{\mathbf{P}}_{2}$ has a shape which resembles number \lq\lq 2\rq\rq. Modes $\overline{\mathbf{P}}_{1}$ and $\overline{\mathbf{P}}_{2}$ are associated with eigenvalues $\xi_A=1$ and $\xi_B=2$, respectively. Finally, at the angular frequency $\omega_3$ we prescribe modes $\overline{\mathbf{P}}_{2}$ and $\overline{\mathbf{P}}_{0}$, associated with eigenvalues $\xi_A=1$ and $\xi_B=2$, respectively. Tables \ref{tab:prescr_modes} and \ref{tab:prescr_modes_list} summarize these choices.

\begin{table}[htb]
    \centering
    \begin{tabular}{c c c}
    \hline
    $i$     & $\omega_i$ (rad/s) & $f_i$ (GHz)\\
    \hline
    $1$ & $12.57 \times 10^9$ & $2.0$ \\
    $2$ & $15.71 \times 10^9$ & $2.5$ \\
    $3$ & $18.85 \times 10^9$ & 3.0 \\
        \hline
    \end{tabular}

    \caption{The prescribed resonance frequencies.}
    \label{tab:TB1}
\end{table}

\begin{table}[htb]
    \centering
    \begin{tabular}{c c c c}
        \hline
        & $\omega_1$ & $\omega_2$ & $\omega_3$  \\
        \hline
        $\xi_A$ & $\mathbf{P}_0$& $\mathbf{P}_1$ & $\mathbf{P}_2$\\
        $\xi_B$ & $\mathbf{P}_1$ & $\mathbf{P}_2$ & $\mathbf{P}_0$\\
        \hline
    \end{tabular}
    \caption{The prescribed modes at each angular frequency, together with the related eigenvalues.}
    \label{tab:prescr_modes}
\end{table}

\begin{table}[htb]
    \centering
    \begin{tabular}{cc}
        \hline
        No. & Mode\\
        \hline
         1 & $(\omega_1, \xi_A, \overline{\mathbf{P}}_{0})$\\
         2 & $(\omega_1, \xi_B, \overline{\mathbf{P}}_{1})$\\
         3 & $(\omega_2, \xi_A, \overline{\mathbf{P}}_{1})$\\
         4 & $(\omega_2, \xi_B, \overline{\mathbf{P}}_{2})$\\
         5 & $(\omega_3, \xi_A, \overline{\mathbf{P}}_{2})$\\
         6 & $(\omega_3, \xi_B, \overline{\mathbf{P}}_{0})$\\
        \hline
    \end{tabular}
    \caption{List of the prescribed modes.}
    \label{tab:prescr_modes_list}
\end{table}

\begin{figure*}
    \centering
    \includegraphics[width=\textwidth]{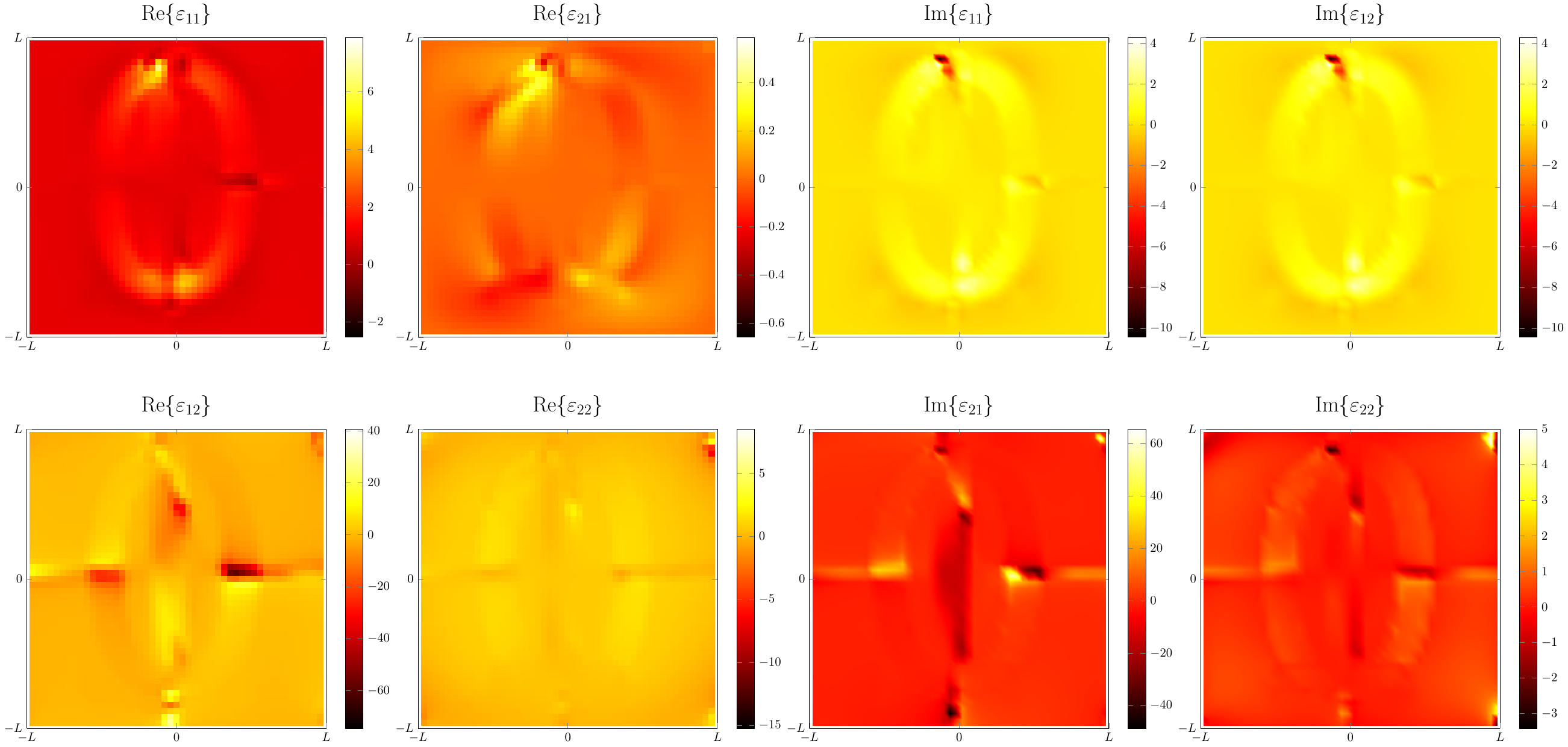}
    \caption{\textbf{Plot of tensor $\varepsilon(\cdot,\omega_1)$.} Spatial profile of the real and imaginary part of the elements of the synthetized permittivity tensor at angular frequency $\omega_1$.}
    \label{fig:EpsOmega1}
\end{figure*}

\begin{figure*}
    \centering
    \includegraphics[width=\textwidth]{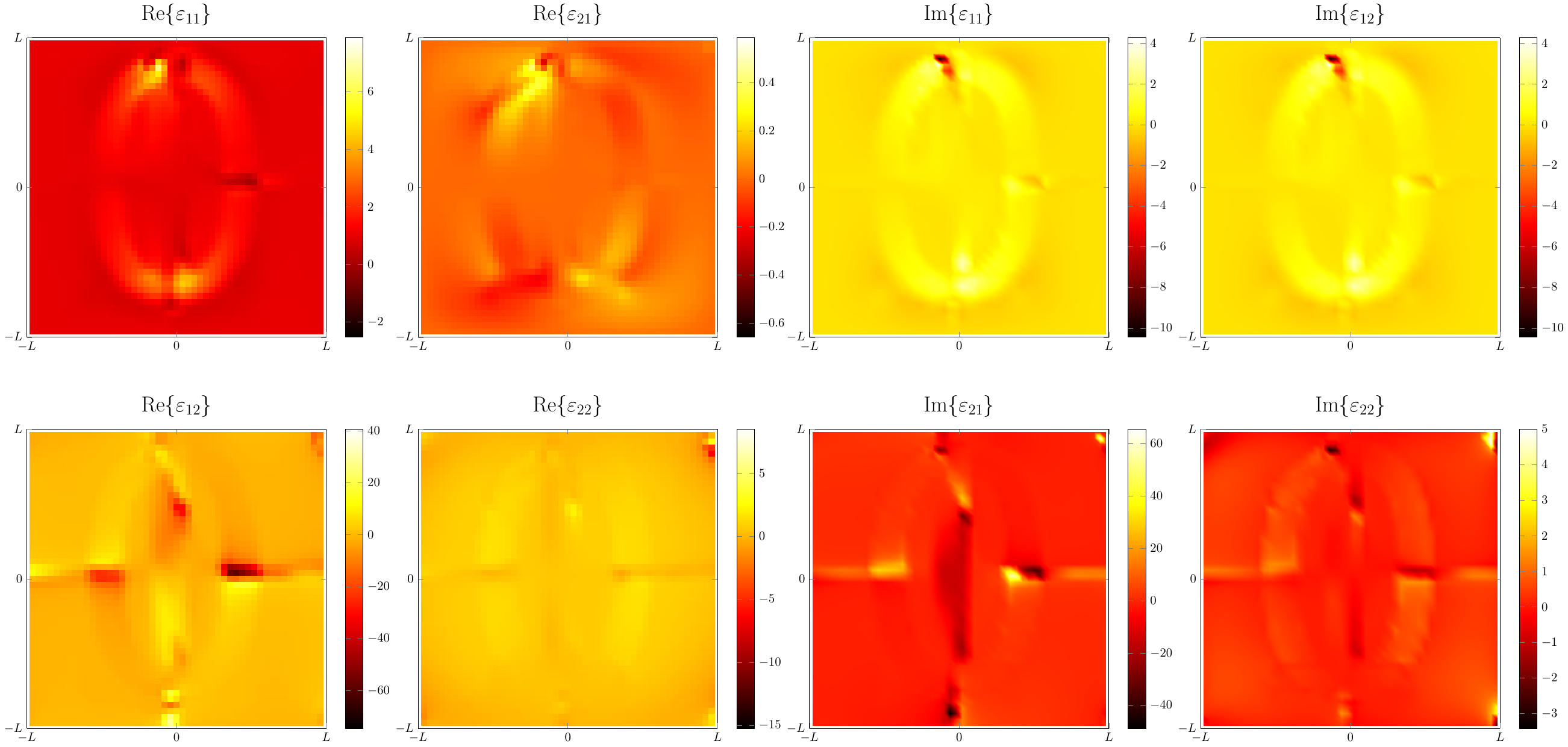}
    \caption{\textbf{Plot of tensor $\varepsilon(\cdot,\omega_2)$.} Spatial profile of the real and imaginary part of the elements of the synthetized permittivity tensor at angular frequency $\omega_2$.}
    \label{fig:EpsOmega2}
\end{figure*}

\begin{figure*}
    \centering
    \includegraphics[width=\textwidth]{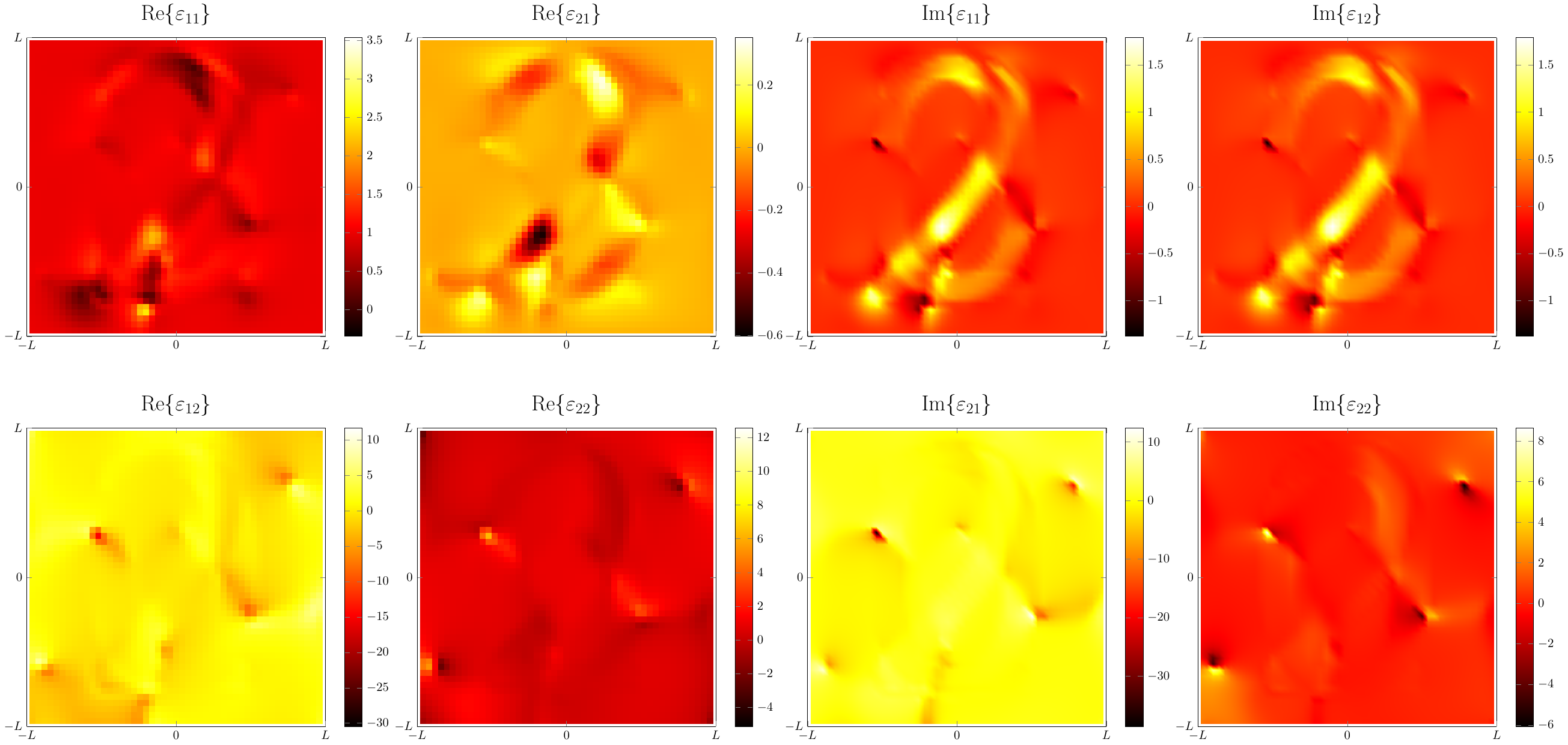}
    \caption{\textbf{Plot of tensor $\varepsilon(\cdot,\omega_3)$.} Spatial profile of the real and imaginary part of the elements of the synthetized permittivity tensor at angular frequency $\omega_3$.}
    \label{fig:EpsOmega3}
\end{figure*}

\begin{figure}
    \centering
    \includegraphics[width=\columnwidth]{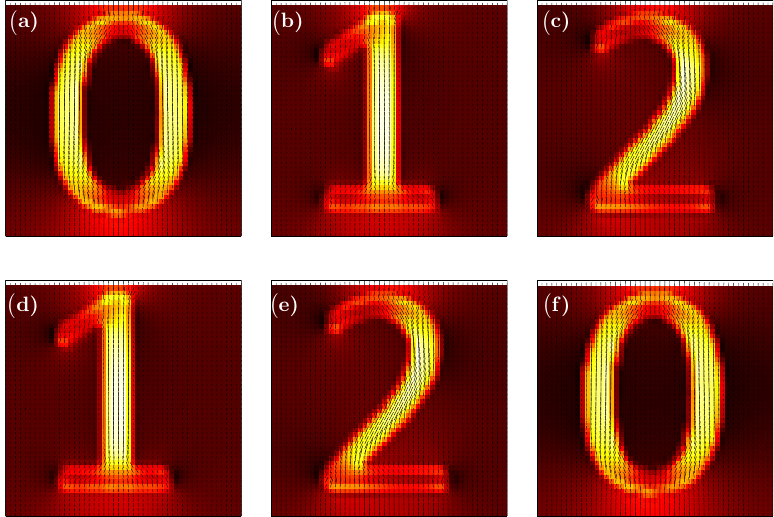}
    \caption{\textbf{Response of the scatterer at each resonant frequency.} Polarization density fields induced in a translational invariant cylinder with square $L \times L$ cross section excited by a linearly polarized plane wave at the frequencies $\omega_{0,1}$ (a), (d); $\omega_{0,2}$ (b), (e); and $\omega_{0,3}$ (c), (f). In panels (a-c) the susceptibility of the dielectric object is $\chi ( \mathbf{r},\omega )$, corresponding to $\xi_\mathtt{A} = 1$. In panels (d-f) the susceptibility is $\chi ( \mathbf{r},\omega ) /2$, corresponding to $\xi_B=2$. Tunability is clearly demonstrated.}
    \label{fig:ScatteringModes}
\end{figure}

The synthesis is carried out in two steps: i) we evaluate $\overleftrightarrow{\gamma}_i \left( \mathbf{r} \right)$ at the three prescribed frequencies; ii) we interpolate the corresponding dielectric susceptibility as in Eq. \eqref{eq:DielectricSusceptibility}, by solving \eqref{eq:PointMatchingA} and \eqref{eq:PointMatchingB}.

In the first step, the theory for the synthesis of two isofrequential modes 
is applied at each individual angular frequency using equation \eqref{eq:solution-two}: (i) for $(\omega_1, \xi_A, \overline{\mathbf{P}}_{0})$ and $(\omega_1, \xi_B, \overline{\mathbf{P}}_{1})$ at $\omega_1$, (ii) for $(\omega_2, \xi_A, \overline{\mathbf{P}}_{1})$ and $(\omega_2, \xi_B, \overline{\mathbf{P}}_{2})$ at $\omega_2$ and (iii) for $(\omega_3, \xi_A, \overline{\mathbf{P}}_{2})$ and $(\omega_3, \xi_B, \overline{\mathbf{P}}_{0})$ at $\omega_3$. 

Figures \ref{fig:EpsOmega1}, \ref{fig:EpsOmega2}, and \ref{fig:EpsOmega3} show the real and imaginary part of every element of the relative dielectric permittivity tensor $\varepsilon_{R,k}=\chi_k+1$, at $\omega_1$, $\omega_2$, and $\omega_3$, respectively.

To validate the proposed method, we performed two tests, where the dielectric susceptibility profile is either $\chi^\mathtt{A} ( \mathbf{r}, \omega ) = \chi ( \mathbf{r}, \omega ) / \xi^\mathtt{A}$ or $\chi^\mathtt{B} ( \mathbf{r}, \omega ) = \chi ( \mathbf{r}, \omega ) / \xi^\mathtt{B}$, where $\chi ( \mathbf{r}, \omega )$ is the outcome of the synthesis of modes.

The first test was a \textit{direct test} and it consisted in i) computing the modes at the three frequencies and in ii) comparing them with the prescribed polarization density field. This test was passed successfully. 


 As second test, we evaluate the induced polarization density fields at the three frequencies $\omega_1$, $\omega_2$, and $\omega_3$, when the cylinder is excited by a linearly polarized plane wave, propagating along the horizontal axis. These polarization fields are showed in Fig. \ref{fig:ScatteringModes} (e-c) assuming a susceptibility tensor $\chi^\mathtt{A}(\mathbf{r},\omega)$ and in Fig. \ref{fig:ScatteringModes} (d-f) for $\chi^\mathtt{B}$.
The induced polarization density fields is very close to the prescribed modes. In quantitative terms, Table \ref{tab:ErrorP}, shows the $2$-norm of the relative difference between the actual $\mathbf{P}$ and its projection along the subspaces generated by the prescribed modes, at each specific angular frequency:
\begin{equation}
\rho_k^i = \frac{\left\Vert \mathbf{P}_i\left(  \cdot,\omega_k\right)  \mathbf{-} \Pi^i_{k} \mathbf{P}_i\left( \cdot, \omega_k\right)  \right\Vert}{\left\Vert
\mathbf{P}_i\left( \cdot, \omega_k \right) \right\Vert }
\label{eq08}
\end{equation}
with $k=1,2,3$ and $i=\mathtt{A},\mathtt{B}$. In \eqref{eq08}, $\mathbf{P}_\mathtt{A} \left(  \cdot,\omega_k\right)$ and $\mathbf{P}_B \left(  \cdot,\omega_k\right)$ are the polarization vectors at $\omega_k$ and for material $\mathtt{A}$ and $\mathtt{B}$, $\Pi^{\mathtt{A}}_{k}$ and $\Pi^{\mathtt{B}}_{k}$ are the projector into the linear space for the modes at the $k-$th angular frequency $\omega_k$ and for material $\mathtt{A}$ and $\mathtt{B}$. The detail about projectors $\Pi^{\mathtt{A}}_{k}$s and $\Pi^{\mathtt{B}}_{k}$s is given in Table \ref{tab:projs}.
\begin{table}[htb]
    \centering
    \begin{tabular}{c c c c}
        \hline
        & $\omega_1$ & $\omega_2$ & $\omega_3$  \\
        \hline
        $A$ & $\spn\left\{  \overline{\mathbf{P}}_0\right\}$ & $\spn\left\{  \overline{\mathbf{P}}_1\right\}$ & $\spn\left\{  \overline{\mathbf{P}}_2\right\}$\\
        $B$ & $\spn\left\{  \overline{\mathbf{P}}_1\right\}$ & $\spn\left\{  \overline{\mathbf{P}}_2\right\}$ & $\spn\left\{  \overline{\mathbf{P}}_0\right\}$ \\
        \hline
    \end{tabular}
    \caption{Description of the vector spaces underlying the projectors. For instance, projector $\Pi_1^{\mathtt{A}}$ projects a polarization density onto $\spn\left\{  \overline{\mathbf{P}}_0\right\}$, or $\Pi_2^{\mathtt{B}}$ projects onto $\spn\left\{  \overline{\mathbf{P}}_2\right\}$.}
    \label{tab:projs}
\end{table}
We stress that
$\mathbf{P}_i \left( \cdot, \omega_k\right)  $ is the polarization vector for the
physical system under the prescribed illumination at $\omega_k$.
\begin{table}[htb]
    \centering
    \begin{tabular}{cccc}
    \hline
    & $2$GHz & $2.5$GHz & $3$GHz \\
    \hline
    Material 1 & $6.0924 \times 10^{-3}$ & $1.8811\times 10^{-2}$ & $1.2875\times 10^{-2}$ \\
    Material 2 & $2.3116\times 10^{-2}$ & $2.2611\times 10^{-2}$ & $5.9386\times 10^{-3}$ \\
    \hline
    \end{tabular}
    \caption{Relative difference between $\mathbf P$ and its projection along the subspaces generated by the prescribed modes, at the frequencies of interest.}
    \label{tab:ErrorP}
\end{table}

This example clearly illustrates the concept of tunability of the resonant response: by just uniformly halving the value of the susceptibility distribution (passing from $\chi^\mathtt{A}$ to $\chi^\mathtt{B}$) the resonance modes in correspondence of the peaks change from the ordered sequence \lq\lq 0\rq\rq, \lq\lq 1\rq\rq, \lq\lq 2\rq\rq to \lq\lq 1\rq\rq, \lq\lq 2\rq\rq, \lq\lq 0\rq\rq.

\begin{figure}
    \centering
    \includegraphics[width=\columnwidth]{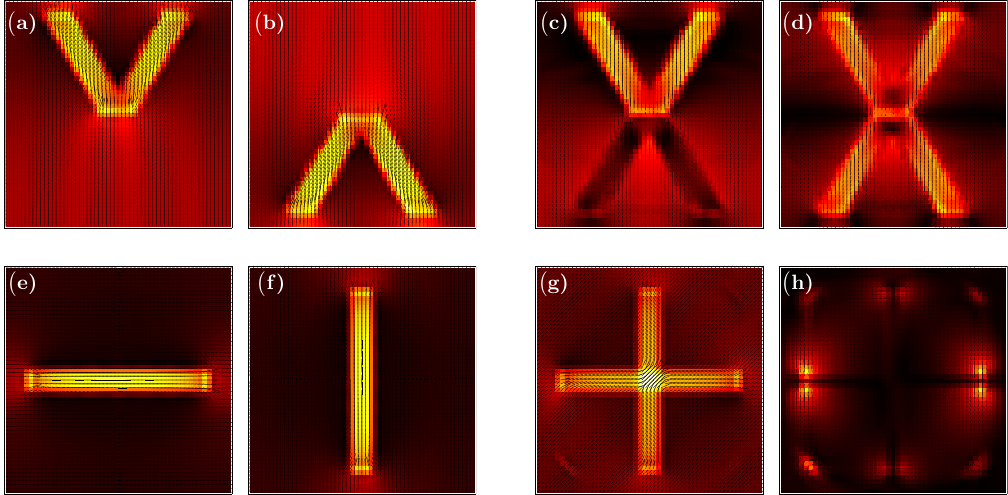}
    \caption{\textbf{Response of the scatterer at each resonant frequency.} Polarization density fields $\mathbf{P}_\land$ (a) and $\mathbf{P}_\vee$ (b) prescribed at $\omega_{0,1}$; and polarization density fields $\mathbf{P}_{-}$ (e) and $\mathbf{P}_|$ (f) prescribed at $\omega_{0,2}$. Real (c) and imaginary (d) part of the polarization density field induced by a plane wave linearly polarized along the diagonal of the (square) cross section at frequency $\omega_{0,1}$. Real (g) and imaginary (h) part of the polarization density field induced at frequency $\omega_{0,2}$. Tunability is clearly demonstrated.}
    \label{fig:ScatteringX}%
\end{figure}

{\it Tunability.} In this second application we determine the dielectric susceptibility by synthesizing at the frequency $\omega_1$ the degenerate modes $\overline{\mathbf{P}}_\land$ and $\overline{\mathbf{P}}_\vee$, whose polarization density field distribution resembles the characters $\land$ and $\vee$, respectively; and at $\omega_2$ the degenerate modes $\overline{\mathbf{P}}_{-}$ and $\overline{\mathbf{P}}_{|}$, whose prescribed field distribution resembles the characters $-$ and $|$, respectively. To validate the performed synthesis, we excite the infinite cylinder with a plane wave polarized along $(\mathbf{e}_1+\mathbf{e}_2)/\sqrt{2}$.  We show the real and imaginary part of the induced polarization field distributions at $\omega_1$ in Figures \ref{fig:ScatteringX}(c), (d), and in Figures \ref{fig:ScatteringX}(g), (h) at $\omega_2$. It is immediately apparent that at $\omega_1$ the induced polarization field is a linear combination of the two prescribed degenerated modes $\overline{\mathbf{P}}_\land$ and $\overline{\mathbf{P}}_\vee$, while at $\omega_1$ the induced polarization field is a linear combination of $\overline{\mathbf{P}}_{-}$ and $\overline{\mathbf{P}}_|$. From the quantitative perspective, the $2-$norm relative difference $\rho$ between the actual $\mathbf{P}$ and its projection along the subspaces generated by the prescribed degenerated modes, is equal at $2.9908 \times 10^{-2}$ at $\omega_1$ and $3.5310 \times 10^{-2}$ at $\omega_2$. In this case $\Pi_1$ projects onto $\spn\left\{  \overline{\mathbf{P}}_\land, \overline{\mathbf{P}}_\vee \right\}$, whereas $\Pi_2$ projects onto $\spn \left\{  \overline{\mathbf{P}}_-, \overline{\mathbf{P}}_| \right\}$.

{\it Essential modes.}
This final application case demonstrates a key feature of the Theory of the Synthesis of Modes, i.e. the concept of \emph{Essential Modes}.
Specifically, given a scatterer operated at a prescribed angular frequency $\omega_1$ and described by the dielectric susceptivity tensor $\chi(\cdot,\omega_1)$, we compute two resonance modes $(\omega_1, \xi_A,\overline{\mathbf{P}}_{A})$ and $(\omega_1, \xi_B,\overline{\mathbf{P}}_{B})$ and, then, we apply our Theory of the Synthesis to these modes. Since the tensor of the dielectric permittivity is in an one-to-one correspondence with two arbitrary modes, as discussed in a previous Section, we expect that the tensor $\chi_s(\cdot,\omega_1)$ of the dielectric permittivity synthesized by means of $(\omega_1, \xi_A,\overline{\mathbf{P}}_{A})$ and $(\omega_1, \xi_B,\overline{\mathbf{P}}_{B})$ via \eqref{eq:solution-two}, is equal to $\chi(\cdot,\omega_1)$.

The scatterer of this example consists of a coated (thickness $100$ nm) circular (radius $200$ nm) gold nanorod operated at $f=500$ THz ($\omega_1=\pi \times 10^{15}$ rad/s, free-space wavelength of $600$ nm). The relative dielectric permittivity of the gold nanoparticle is $9.44-j 1.51$, whereas that of the coating is $4$.

\begin{figure}
    \centering
    \includegraphics[width=\columnwidth]{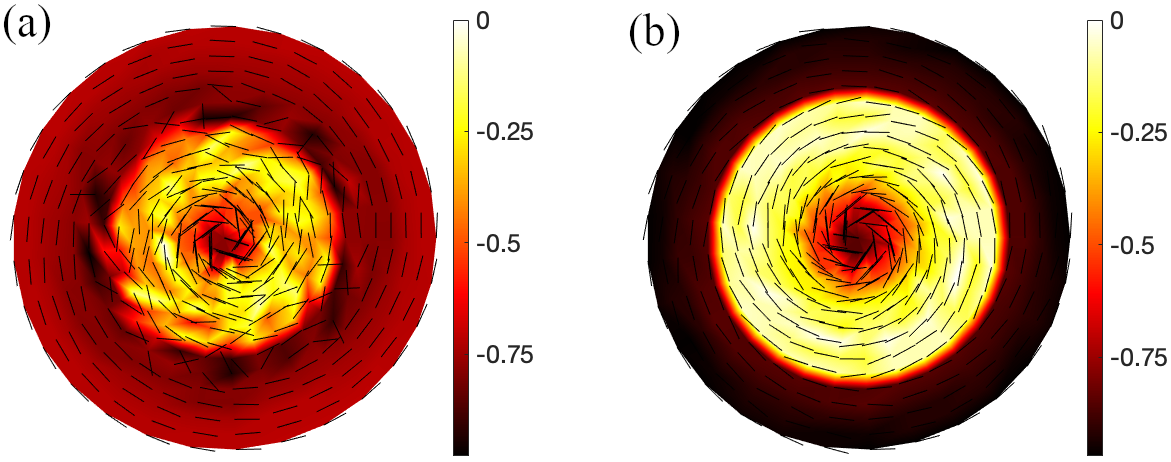}
    \caption{\textbf{Polarization density field for a resonant mode.} Plot of the real (left) and imaginary (right) parts of $\overline{\mathbf{P}}_{A}$ at $f=500$ THz. The corresponding eigenvalue is \textcolor{black}{$\xi_A=-22.29+j16.09$}.}
\label{fig:PA}
\end{figure}

\begin{figure}
    \centering
    \includegraphics[width=\columnwidth]{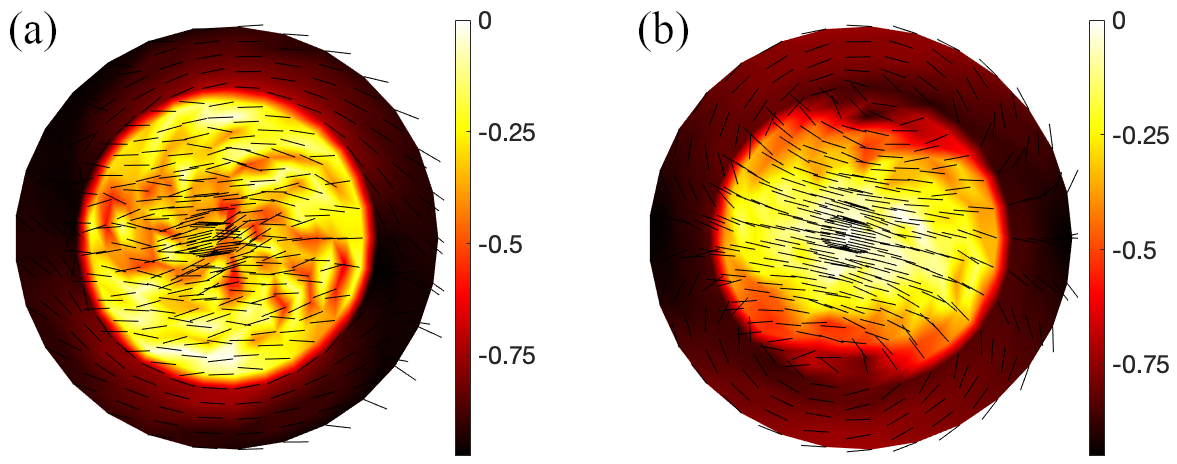}
    \caption{\textbf{Polarization density field for a resonant mode.} Plot of the real (left) and imaginary (right) parts of $\overline{\mathbf{P}}_{B}$ at $f=500$ THz. The corresponding eigenvalue is \textcolor{black}{$\xi_B=13.82+j16.07$}.}    
\label{fig:PB}
\end{figure}

Figures \ref{fig:PA} and \ref{fig:PB} show the real and imaginary parts for the selected modes $\overline{\mathbf{P}}_{A}$ and $\overline{\mathbf{P}}_{B}$. The synthesized dielectric permittivity tensor is almost equal to that of the prescribed scatterer. As a figure of merit we evaluated the maximum relative error over the scatterer domain $\Omega$:
\begin{equation}
\label{eq:err}
    e=\max_{\mathbf{r} \in \Omega} \frac{||\chi(\mathbf{r},\omega_1)-\chi_s(\mathbf{r},\omega_1)||_2}{||\chi(\mathbf{r},\omega_1)||_2},
\end{equation}
which, in this case, is equal to $3.3 \times 10^{-11}$. In \eqref{eq:err} $\chi$ is the prescribed tensor of the dielectric susceptibility, whereas $\chi_s$ is the tensor of the synthesized dielectric susceptibility.

\section*{Conclusions}

In this work we introduced a theoretical framework to find the permittivity profile of a dielectric object to synthesize at will its resonant modes. Specifically, we are able to control the spatial distribution of the polarization density field  and the resonance frequency of a set of modes. The equations for the synthesis are straightforward and in an explicit form, making them suitable for specific customization. Moreover, we can prescribe the modes at many different frequencies.

The only limit, arising from the underlying physics, consists in the possibility of assigning at most two modes to each individual frequency and eigenvalue (up to three modes in a 3D setting). Indeed, from the theory of the synthesis of modes arises naturally that, at a prescribed angular frequency, the dielectric susceptivity tensor is in one-to-one correspondence with two of its modes, that we termed as \emph{essential modes}.

We also demonstrated the concept of \emph{tunability}: the proposed approach enables the design of the permittivity of a dielectric object that not only allows the synthesis at will of its resonant modes, but also allows to changes the resonant modes of the dielectric object in a controlled manner, by multiplying the designed permittivity by a proper multiplicative factor.

With this theoretical framework, future development will be aimed to design a real world material approximating the synthesized dielectric susceptibility. Metamaterials are the natural candidates to this purpose.

{
The method introduced can be transplanted to different linear physical  systems, where the constitutive relationship is linear and local, including thermal and mechanical systems.}

\section*{Methods}
All the numerical calculations have been carried out by using the numerical method of \cite{richmond_te-wave_1966}. All the value of the parameters used for generating numerical results have been included into the article.

\section*{Data Availability}
All the data supporting the conclusions of this study are included in the
article. Source data are provided with this paper.

\section*{Code Availability}
The computer code and algorithm that support the findings of this
study are available from the corresponding author on request.

\appendix
\section{Green Function}

\label{sec:Green}
The component of the Green function for the \texttt{TEz} illumination are
\begin{align*}
G_{11}\left(  \mathbf{r}\right)   & =-\frac{\zeta_{0}}{4r^{3}}\left[
krx_{2}^{2}H_{0}\left(  kr\right)  +\left(  x_{1}^{2}-x_{2}^{2}\right)
H_{1}\left(  kr\right)  \right] \\
G_{12}\left(  \mathbf{r}\right)   & =-\frac{\zeta_{0}}{4r^{3}}x_{1}
x_{2}\left[  2H_{1}\left(  kr\right)  -krH_{0}\left(  kr\right)  \right] \\
G_{21}\left(  \mathbf{r}\right)   & =G_{12}\left(  \mathbf{r}\right) \\
G_{22}\left(  \mathbf{r}\right)   & =-\frac{\zeta_{0}}{4r^{3}}\left[
krx_{1}^{2}H_{0}\left(  kr\right)  +\left(  x_{2}^{2}-x_{1}^{2}\right)
H_{1}\left(  kr\right)  \right]  ,
\end{align*}
being $\zeta_{0}$ the characteristic impedance of vacuum, $k=\omega/c_{0}$ the
wavenumber, and $c_{0}$ the speed of light in vacuum.

\section{Lorentz-Drude expansion function}

\label{sec:LorentzDrude}
\begin{figure}
    \centering
    \includegraphics[width=\columnwidth]{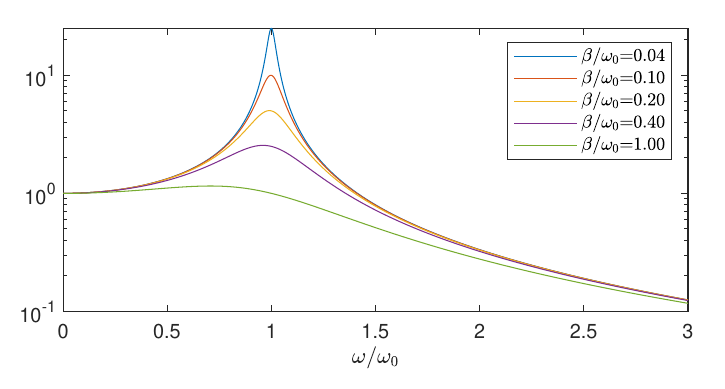}
    \caption{\textbf{The elementary Lorentz-Drude function.} Plot of the normalized amplitude of the Lorentz-Drude expansion functions, at different $\beta / \omega_0$ ratios.}
    \label{fig:NormPlot}
\end{figure}

The (normalized) amplitude of the elementary Lorentz-Drude expansion function is:
\begin{equation}
\frac{\left| \varphi (\omega) \right|}{\left( \omega_{p} / \omega_{0} \right)^{2}} =\frac{1}{\sqrt{\left[ 1 - \left( \frac{\omega}{\omega_0} \right)^{2} \right]^2  +\left( \frac{\omega}{\omega_0} \right)^2 \left( \frac{\beta}{\omega_0} \right)^2}}.
\label{eq:DrudeAmpl}
\end{equation}
Its maximum value is
\begin{equation}
\frac{\left| \varphi (\omega) \right|_{max}}{\left( \omega_{p} / \omega_{0} \right)^{2}} =\frac{1}{\frac{\beta}{\omega_0}\sqrt{1 + \frac{3}{4} \left( \frac{\beta}{\omega_0} \right)}}.
\end{equation}
and it is achieved at
\begin{equation}
    \frac{\omega}{\omega_0} = \sqrt{1+\frac{1}{2} \left( \frac{\beta}{\omega_0} \right)^2}
\end{equation}
The plot of \eqref{eq:DrudeAmpl} for different $\beta / \omega_0$ ratios is showed in Figure \ref{fig:NormPlot}.

\end{document}